\documentclass[reprint, amsmath,amssymb, aps, superscriptaddress,floatfix]{revtex4-2}
\usepackage{float}
\usepackage{graphicx}% Include figure files
\usepackage{dcolumn}% Align table columns on decimal point
\usepackage{bm}% bold math
\usepackage[colorlinks = true, linkcolor = blue,  urlcolor  = blue, citecolor = magenta, anchorcolor = blue]{hyperref}
\usepackage{color}
\begin{document}

\title{Structure and dynamics of motor-driven microtubule bundles}
\author{Bezia Lemma}
\thanks{These authors contributed equally}
\affiliation{Physics Department, Harvard University, Cambridge, MA 02138, USA}
\affiliation{Physics Department, Brandeis University, Waltham, MA 02453, USA}
\affiliation{Physics Department, University of California, Santa Barbara, CA 93106, USA}

\author{Linnea Lemma}
\thanks{These authors contributed equally}
\affiliation{Physics Department, Brandeis University, Waltham, MA 02453, USA}
\affiliation{Physics Department, University of California, Santa Barbara, CA 93106, USA}

\author{Stephanie C. Ems-McClung}
\affiliation{Medical Sciences, Indiana University School of Medicine, Bloomington, IN 47405, USA}
    
\author{Claire E. Walczak}
\affiliation{Medical Sciences, Indiana University School of Medicine, Bloomington, IN 47405, USA}

\author{Zvonimir Dogic}
\affiliation{Physics Department, University of California, Santa Barbara, CA 93106, USA}
\affiliation{Physics Department, Brandeis University, Waltham, MA 02453, USA}
\affiliation{Biomolecular Science \& Engineering Department, University of California, Santa Barbara, CA 93106, USA}

\author{Daniel J. Needleman}
\affiliation{John A. Paulson School of Engineering and Applied Sciences, Harvard University, Cambridge, MA 02138, USA}
\affiliation{Molecular \& Cellular Biology Department, Harvard University, Cambridge, MA 02138, USA}
\affiliation{Center for Computational Biology, Flatiron Institute, New York, NY 10010}

\date{\today}

\begin{abstract}
Connecting the large-scale emergent behaviors of active materials to the microscopic properties of their constituents is a challenge due to a lack of data on the multiscale dynamics and structure of such systems. We approach this problem by studying the impact of polyethylene glycol, a crowding agent, on bundles of microtubules and kinesin-14 molecular motors. Bundles assembled in the presence of either low or high concentrations of polyethylene glycol generate similar net extensile behaviors. However, as polyethylene glycol concentration is increased, the motion of microtubules in the bundles transition from bi-directional sliding with extension to pure extension with no sliding. Small-angle X-ray scattering shows that the transition in microtubule dynamics is concomitant with a rearrangement of microtubules in the bundles from an open hexagonal to a compressed rectangular lattice. These results demonstrate that bundles of microtubules and molecular motors can display similar mesoscopic extensile behaviors despite having very different internal structures and dynamics.
\end{abstract}

\maketitle
%\section{\label{sec:intro}Introduction}

Kinesin molecular motors hydrolyze chemical energy to generate pico-Newton forces and step along filamentous microtubules \cite{yildiz2008intramolecular, visscher1999single}. Several kinesin motors and synthetic multi-motor clusters can simultaneously bind adjacent microtubules and induce their relative sliding. The motion of such elemental building blocks can cascade across several levels of hierarchy to generate diverse internally driven dynamics ranging from localized and bulk contractions to turbulent-like flows, and active foams~\cite{nedelec1997,hentrich2010microtubule,tim2012,peter2017,senoussi2019tunable,gagnon2020shear,lemma2022active}. Quantitatively connecting the microscopic behaviors of molecular motors and filaments to large-scale emergent properties remains a challenge in the field of active matter~\cite{murrell2012f,lenz2012contractile,gao2015,shelley2016rev,belmonte2017theory,belmonte2017theory,yang2021physical,vliegenthart2020filamentous,sebastian2019,lenz2020reversal,yan2022toward,najma2021dual}. 

Experiments measuring the relative speed between two microtubules within a single bundle or a nematic domain have revealed two distinct types of microscopic dynamics. Two-dimensional active nematics driven by kinesin-1 clusters displayed a broad unimodal distribution of relative velocities. These microtubule velocities were significantly slower than those of unloaded motors~\cite{lemma2021multiscale}. In comparison, bundles of aligned microtubules driven by kinesin-14 displayed a bi-modal sharply peaked distribution of velocities, with antiparallel populations of microtubules sliding apart at the speed of the unloaded molecular motor~\cite{sebastian2019}.

Intriguingly, both systems can produce extensile stresses, generating turbulent-like flows. In addition to being powered by different kinesin motors, the two systems differ by the presence of polyethylene glycol (PEG) in kinesin-1 active-nematics and its absence in kinesin-14 bundles. PEG is a crowding/depletion agent that induces an effective attraction between microtubules and can even alter their structure \cite{asakura1954interaction,dan2004,feodor2015}.

To gain insight into the microscale dynamics of active bundles, we studied how PEG concentration influences the internal dynamics and microscopic structure of extensile microtubule bundles powered by kinesin-14. We found that increasing the concentration of PEG induces a transition in the motions of microtubules inside the bundles. Small-Angle X-ray Scattering (SAXS) of microtubule bundles demonstrates that the dynamical transition in the microscopic filament motion is correlated with a structural transition from open hexagonal to tight rectangular packing.

%\section {Results}

\begin{figure*}  
\includegraphics[width=\textwidth]{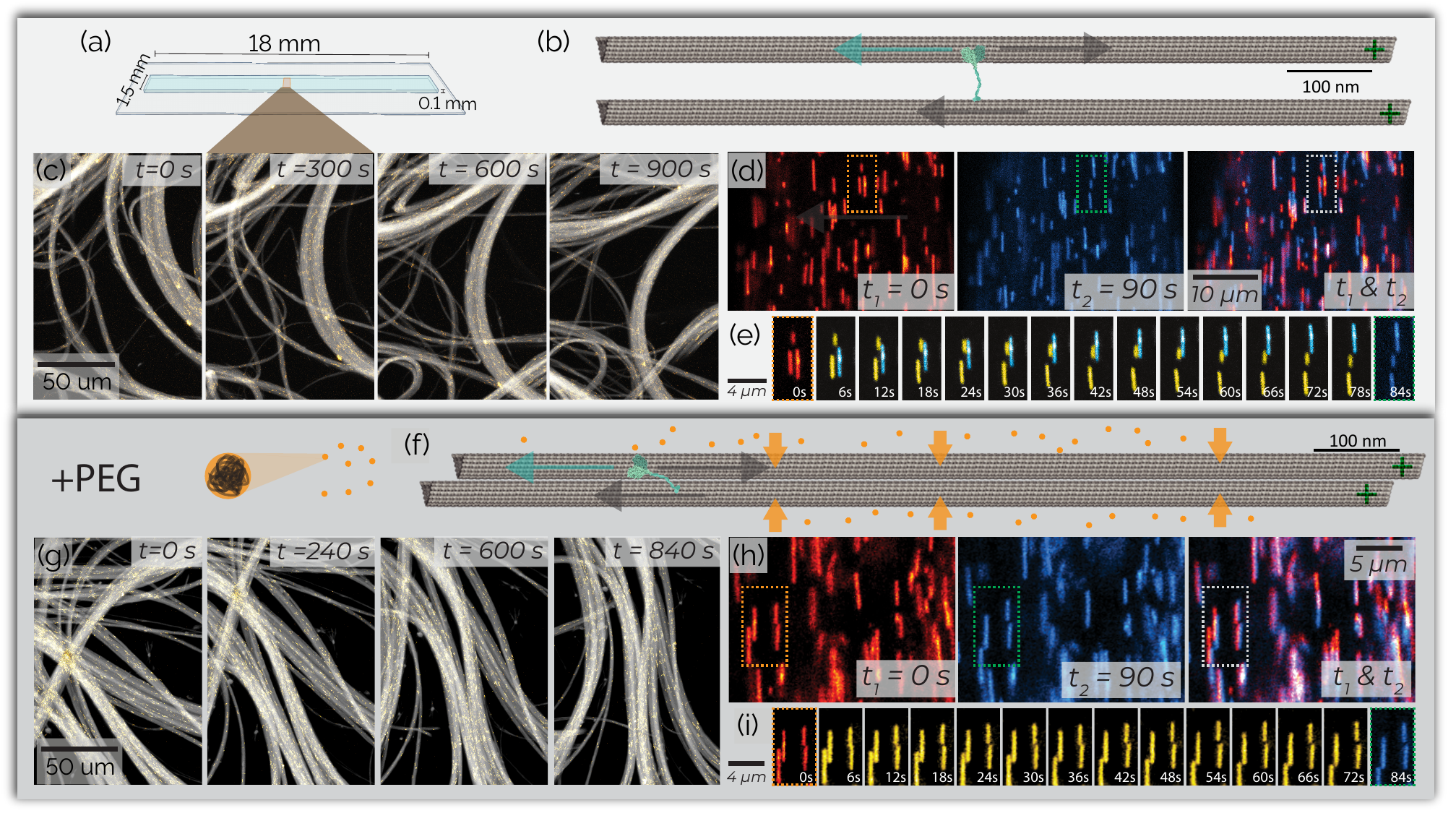}
\caption{ 
\textbf{PEG alters the microtubule motions within extensile bundles formed with kinesin-14.} 
\textbf{(a-b)} Microtubule bundles powered by kinesin-14, without PEG, in a thin chamber. A cyan arrow indicates kinesin motion, and a grey arrow indicates microtubule motion.
\textbf{(c)} A time sequence of a PEG-free kinesin-14/microtubule bundle extending and buckling. 
\textbf{(d)} Time sequence of an extending bundle with yellow tracer microtubules. The initial frame is in red, and the final frame is in blue. The overlay shows the net movement of the microtubules over 90~s. 
\textbf{(e)} Microscale dynamics of tracer microtubules within a bundle without PEG display a smooth sliding motion with distinct populations of downward moving (yellow) and upward moving microtubules (cyan). 
\textbf{(f)} Microtubule bundles with kinesin-14 and PEG. PEG induces attractive forces between microtubules. 
\textbf{(g)} Time sequence of extensile bundles with 1\% PEG and microtubule tracers (yellow). 
\textbf{(h)} Time sequence of an extending bundle with yellow tracer microtubules. 
\textbf{(i)} Microscale dynamics of tracer microtubules within a bundle in the presence of 1\% PEG exhibit a stuttering motion. }
\label{fig:tracers} 
\end{figure*} 

We studied the dynamics of microtubule sliding in bundles driven by kinesin-14 (XCTK2). Kinesin-14 is a minus-end directed dimeric molecular motor that contains a C-terminus motor domain and a passive microtubule-binding N-terminus \cite{claire1997, hallen2011two}. Kinesin-14 crosslinks microtubules and induces their relative sliding~\cite{ems2020rangtp, cai2009}. When we combined kinesin-14 (200 nM) with GMPcPP stabilized microtubules (16 $\mu$M), the microtubules coarsened into bundles [Fig.~\ref{fig:tracers}(a-c)]. These bundles continuously extended, buckled, and annealed [Fig.~\ref{fig:tracers}(c), Vid. 1]. Similar extensile dynamics have previously been observed in microtubule bundles formed with kinesin-4 or various kinesin-1 constructs ~\cite{chandrakar2022engineering,lemma2022active}. However, these other systems contained bundling agents in addition to molecular motors. In contrast, kinesin-14 can yield microtubule bundles without a supplemental bundling agent.

To visualize the motion of individual filaments, we created bundles with kinesin-14 (and no PEG) in which 1 in 10,000 microtubules were labeled [Fig.~\ref{fig:tracers}(d-e)]. Within a bundle, individual microtubules exhibited pronounced and sustained motion along the bundle's long axis. Furthermore, individual filaments hardly ever changed direction, and the bundle exhibited net extensile motion [Fig.~\ref{fig:tracers}(e), Vid. 2]. We next made kinesin-14 bundles with the addition of 1\% w/w PEG and found a similar net bundle extension [Fig.~\ref{fig:tracers}(g), Vid. 1]. However, in the presence of PEG, individual microtubules exhibited markedly different dynamics with reduced speeds and a stuttering motion in which microtubules changed direction [Fig.~\ref{fig:tracers}(f-i), Vid. 2]. Thus, while  both systems produced similar net bundle extension, the microscopic filament dynamics were quite distinct, with and without PEG. 

\begin{figure*} 
\includegraphics[width=\textwidth]{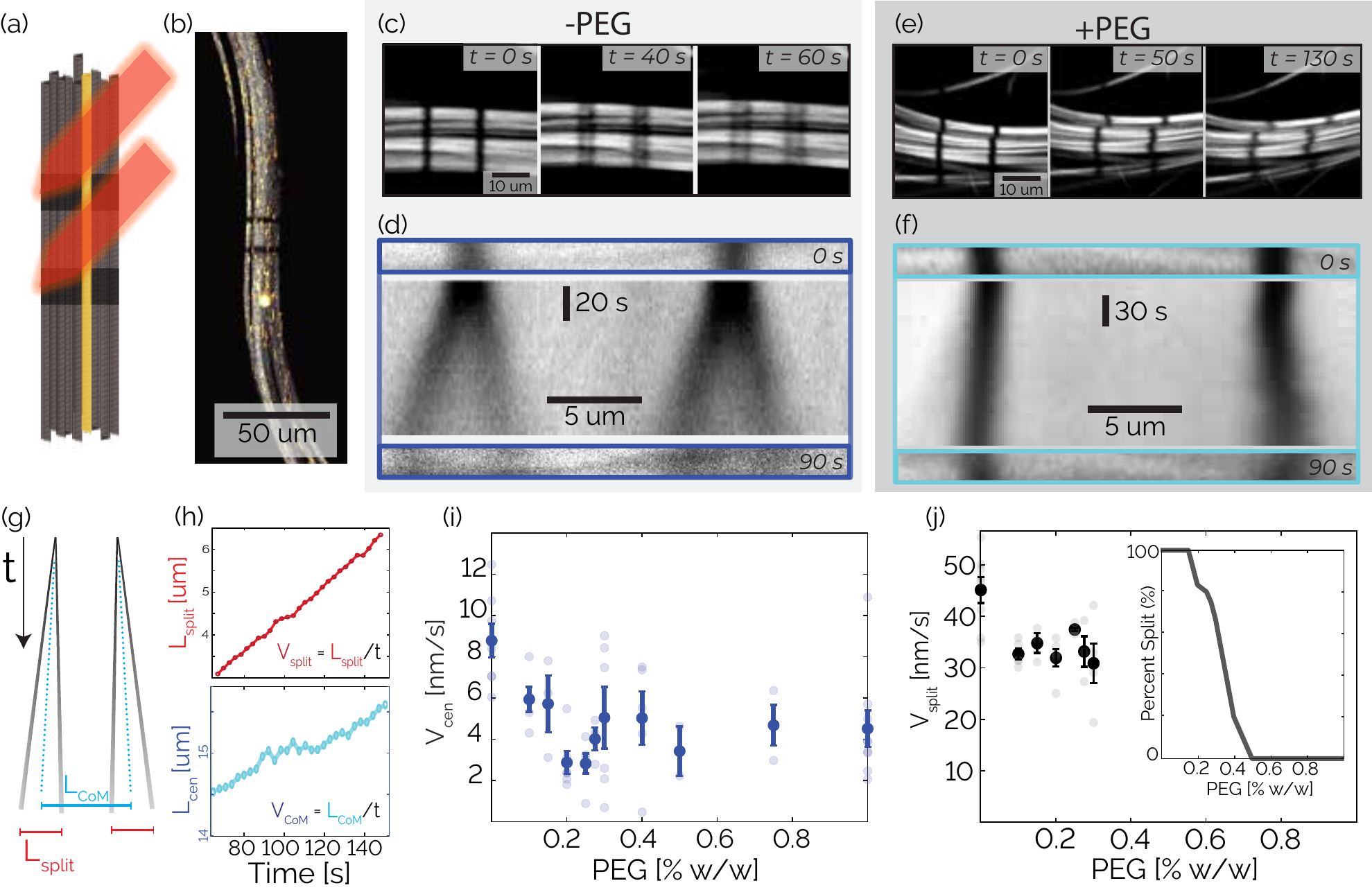}
\caption{ 
\textbf{PEG induces a transition from splitting to non-splitting dynamics regions.}
\textbf{(a)} A laser (orange) bleaches two lines of Alexa-647 labeled microtubules while Azide-DBCO-488 labeled tracer microtubules remain fluorescent. 
\textbf{(b)} A microtubule bundle with tracer microtubules and photobleached lines.
\textbf{(c)} A time sequence of a PEG-free sample in which both bleached regions split into two lines that moved in opposite directions. 
\textbf{(d)} Top: Initial bleach lines. Middle: A space-time image showing the temporal evolution of the bleached lines. Bottom: Bleached patterns at $t=90$ s.
\textbf{(e,f)} A time sequence and space-time image of bleach patterns in the presence of 1\% PEG.
\textbf{(g)} A schematic space-time diagram of splitting bleach lines. The black lines indicate the bleach mark trajectories. Dotted blue lines represent the center of mass of a bleach mark. $L_\textrm{cen}$ is the distance between the center of masses. $L_\textrm{split}$ is the average distance between splitting marks. If the lines do not split  $L_\textrm{split} = 0$.
\textbf{(h)} Experimental data of $L_\textrm{cen}$ and $L_\textrm{split}$ at 0\% PEG, from which $V_\textrm{cen}$ and $V_\textrm{split}$ are calculated.
\textbf{(i)} $V_\textrm{cen}$ as a function of PEG. Faded dots are the results from individual samples; error bars indicate standard error. 
\textbf{(j)} Dark black dots show the average $V_\textrm{split}$, bars show standard error, and faded dots represent individual samples.
Inset: The fraction of bundles that display splitting behavior as a function of \% PEG.}
\label{fig:bleaching} 
\end{figure*}

%\subsection{Microscale filament dynamics from bleaching}
To quantify the dynamics of these two regimes, we photo-bleached two stripes several microns apart and perpendicular to the bundle's long axis [Fig.~\ref{fig:bleaching}(a-b), Vid. 3]. In a sample with 0\% PEG, each bleached region split into two lines that moved in opposite directions along the bundle's long axis [Fig.~\ref{fig:bleaching}(c-d)]. These data are consistent with the observation of sparsely labeled microtubules sliding continuously without changing direction [Fig.~\ref{fig:tracers}]. In contrast, the two bleached regions in 1\% PEG bundles slowly moved apart from each other without splitting [Fig.~\ref{fig:bleaching}(e-f)]. 

We next analyzed space-time plots (i.e. kymographs) to quantify the dynamics of the bleach patterns [Fig.~\ref{fig:bleaching}(c-f)]. We defined $L_{\textrm{split}}$ as the distance between the two peaks which evolve from each bleach mark, while $L_{\textrm{cen}}$ is the distance between the center of masses of the two different bleach marks [Fig.~\ref{fig:bleaching}(g)]. In the cases where the bleach marks split, we calculated the center of mass by weighing the two splitting marks $I_1$ and $I_2$ by their intensity such that $L_\textrm{cen} = \frac{I_1 L_1 + I_2 L_2}{I_1 + I_2}$. On the timescales probed, both $L_{\textrm{split}}$ and $L_{\textrm{cen}}$ increased linearly in time [Fig.~\ref{fig:bleaching}(h), top]. Thus, we calculated $V_\textrm{split}$, which is the speed at which microtubules in one marked region slide relative to their neighbors, and $V_\textrm{cen}$, which is the extension speed of the bundle between the two marked regions for a given lateral spacing [Fig.~\ref{fig:bleaching}(h), bottom]. 

To quantify the transition between the splitting and blurring regimes, we made samples with a range of PEG concentrations. For all samples, we measured $L_\textrm{cen}$, and where possible, $L_\textrm{split}$. The average extensile speed of the system $V_\textrm{cen}$ gradually decreased from $\sim$8.8 nm/s at 0\% PEG to $\sim$4.5 nm/s at 1\% PEG [Fig.~\ref{fig:bleaching}(i)]. The splitting speed  $V_\textrm{split}$ ranged between $\sim$45 nm/s at 0\% PEG and $\sim$30 nm/s at 0.3\% PEG [Fig.~\ref{fig:bleaching}(j)]. Between 0.15\% and 0.4\% PEG, only a fraction of microtubule bundles within a sample displayed splitting of the bleached lines. At 0.3\% PEG, roughly 2/3 of bundles had splitting dynamics, while at 0.4\% PEG, only 1/5 of bundles had splitting dynamics [Fig.~\ref{fig:bleaching}(j), inset]. At these intermediate PEG concentrations, some bleach lines partially split while leaving behind a third fainter line, suggesting a coexistence of the two dynamic regimes [Vid. 3]. 

%\subsection{Small Angle X-Ray Scattering reveals bundle microstructure}

\begin{figure}
\includegraphics[width=\columnwidth]{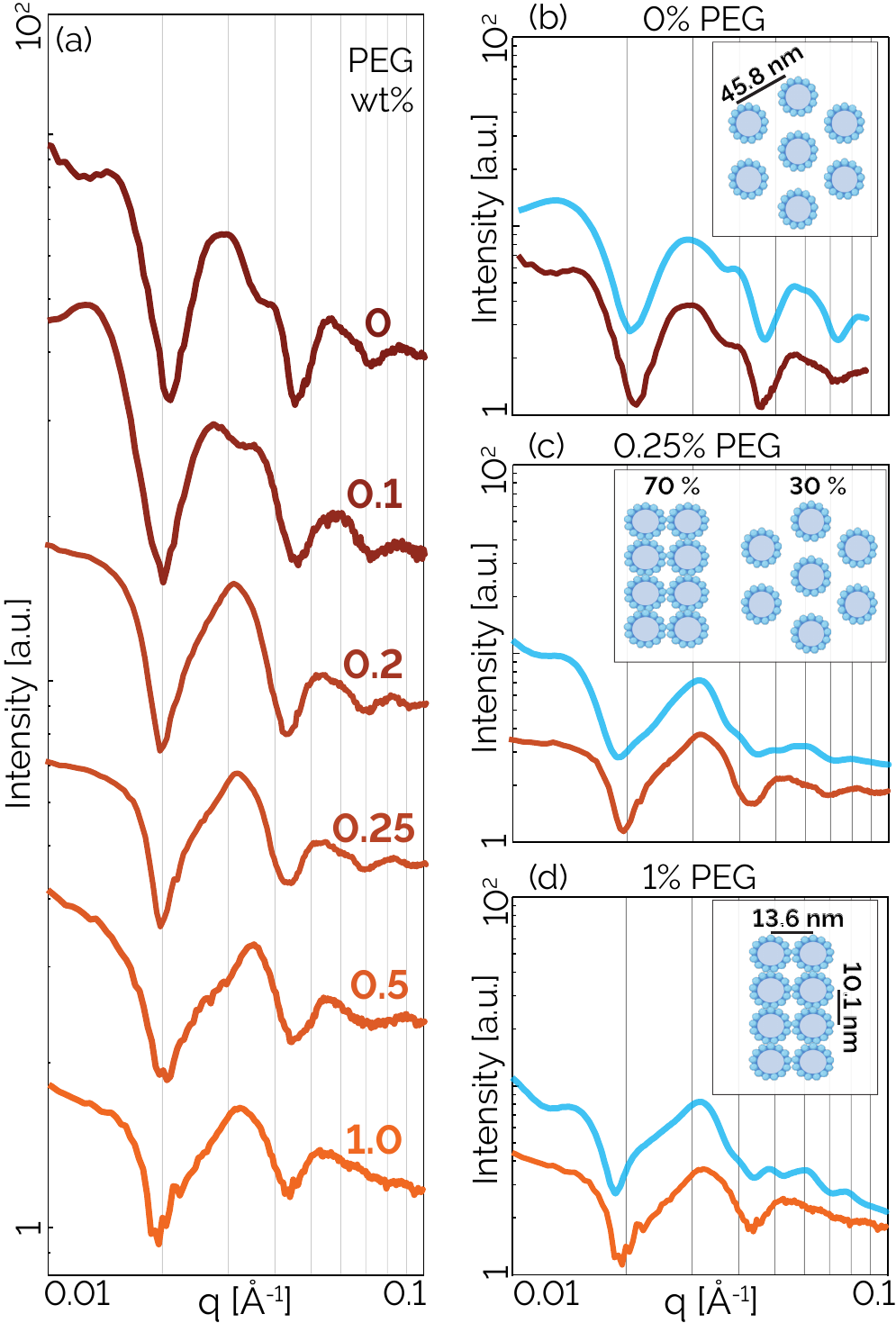}
\caption{\label{fig:SAXS}
\textbf{PEG induces a structural transition in the microtubule bundles.}
\textbf{(a)} Radially averaged SAXS curves of microtubule bundles in the presence of kinesin-14 with increasing PEG concentrations. 
\textbf{(b)} At 0\% PEG, the SAXS curve is consistent with a hexagonal lattice of microtubules with a spacing of $L_h=$ 45.8 nm (cyan).
Data is shifted along the Y-axis for clarity. 
\textbf{(c)} Models of a combination of the hexagonal lattice and the rectangular lattice fit SAXS curves at 0.25\% PEG.
\textbf{(d)} SAXS curves at 1\% PEG match a model (cyan) of microtubules with an ellipsoidal cross-section arranged in a rectangular lattice with $L_a=$ 10.1 nm and $L_b=$ 13.6 nm.}
\end{figure}

Bleaching experiments demonstrate the existence of two dynamical regimes with changing PEG concentration. To gain further insight into the differences between these regimes, we next used Small Angle X-ray Scattering (SAXS) to measure the structure of the bundles. The SAXS scattering curves of samples with and without PEG were notably different, suggesting that the introduction of PEG changed the organization of microtubules in the bundles [Fig.~\ref{fig:SAXS}(a)].

To understand the change of the SAXS curves, we modeled the scattering amplitude of various microtubule bundle packing configurations using a numerical solver that computes X-ray scattering curves of basic filamentous structures distributed in random orientations \cite{DPlus}. We generated several microtubule packing configurations with variable spacing and microtubule number (SI). At 0\% PEG, in the presence of kinesin-14, we found that models of hexagonal lattices with a center-to-center spacing of $L_h = 45.8\pm5$ nm closely match the experimental scattering curve [Fig.~\ref{fig:SAXS}(b)]. In comparison, at 1\% PEG, in the presence of kinesin-14, the scattering curves are consistent with a model of microtubules with an elliptical cross-section close packing in a rectangular lattice ($L_a$ = 13.6 nm, $L_b$ = 10.1 nm) [Fig.~\ref{fig:SAXS}(d)]. The SAXS curves between the 0\% PEG and 1\% PEG concentrations can be modeled as a coexistence of the hexagonal and rectangular lattice patterns  [Fig.~\ref{fig:SAXS}(c)]. The anisotropic compression of the microtubule and the formation of a rectangular lattice have been observed previously in passive bundles~\cite{dan2005}. The matching model uses a small number ($<10$) of microtubules in both rectangular and hexagonal phases. The broad shape of the peaks and the correspondingly small number of microtubules in the lattice models suggest that the filament packing has short-range order at both concentrations.

%\section{\label{sec:discussion} Discussion}

\begin{figure*} 
\includegraphics[width=\textwidth]{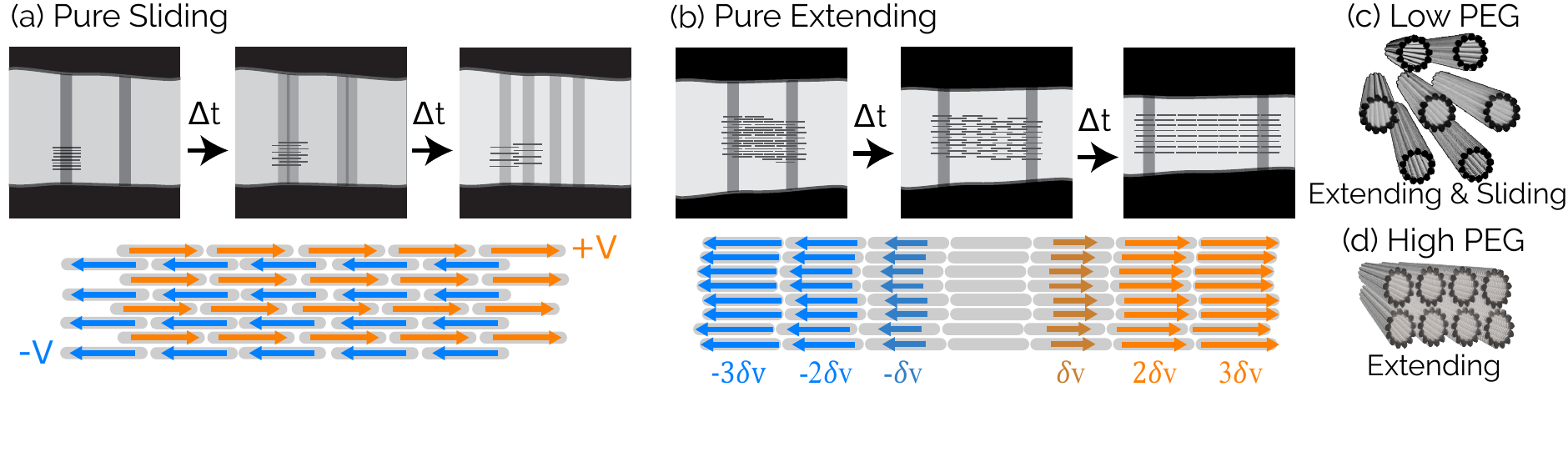}
\caption{ 
\textbf{The transition between splitting and extending microtubule bundles corresponds to a transition from hexagonal to rectangular bundle structure}
\textbf{(a)} An idealized model of pure sliding, in which half the microtubules move with velocity $V$ and half with velocity $-V$. Bleached lines split into two and move at constant velocities, but the center of mass of the two bleached regions remains at a fixed distance. 
\textbf{(b)} An idealized model of pure extension generated by a telescoping elongation of the bundle, which leads to a thinning in the lateral direction. Bleached lines do not split, but their separation increases over time.
\textbf{(c)} In bundles with kinesin-14 and low (or no) PEG, microtubules form a loosely packed hexagonal lattice and exhibit both sliding and extension. 
\textbf{(d)} In bundles with kinesin-14 and high PEG, microtubules form a tightly packed rectangular lattice and exhibit extension with no additional sliding. 
}
\label{fig:diagram} 
\end{figure*} 

In this work, we showed that motor-driven microtubule bundles can exhibit similar large-scale extension despite having very different internal structures and dynamics. Our experiments also highlight the difference between filament sliding and bundle extension. Hypothetically, it is possible to have a bundle in which filaments slide apart from each other without generating any material extension within the bundle. In such a bundle, half the filaments move leftward (velocity $-V$) and the other half rightward (velocity $+V$). As a consequence the two bleach lines  would each split in two (with  $V_\textrm{split} = 2V$) but their centers of mass would remain constant ($V_\textrm{CoM} = 0$) [Fig.~\ref{fig:diagram}(a)]. In contrast, bleaching two lines in a purely extending bundle, which had no additional filament sliding, would result in the relative motion of the center of masses of the marked regions ($V_\textrm{CoM} \neq 0$) with no line splitting  [Fig.~\ref{fig:diagram}(b)]. The velocities of filaments in such a bundle would be telescoping exponentially outwards from all locations. 

Without PEG, the microtubule bundles display both sliding and extending dynamics. This regime has a local hexagonal packing structure with significant space in between the microtubules that are roughly comparable to the kinesin size [Fig.~\ref{fig:diagram}(c)]. Above a threshold PEG concentration, the packing structure becomes rectangular with no significant space between the microtubules. The system exhibits extensile dynamics but with no further sliding [Fig.~\ref{fig:diagram}(d)]. This exstensile dynamics, with no additional sliding, is reminiscent of observations in dense 2D active nematics driven by streptavidin-kinesin-1 clusters \cite{lemma2021multiscale}. These 2D active nematics experiments explicitly showed that the extension velocity increases exponentially along the director. A similar exponential extension presumably exists in the kinesin-14 bundles studied here even though the dynamics appeared linear [Fig.~\ref{fig:bleaching}(h)], likely because an exponential measured over a short time is well approximated as being linear.

Understanding the large-scale dynamics of cytoskeletal active matter has emerged as an important and extensively studied research theme. In comparison, relatively little is known about the microscopic structure of basic units that power these large-scale non-equilibrium behaviors. Our work provides structural insight into extensile microtubule bundles, which drive diverse forms of microtubule-based active matter. However, this work is only the first step, as the microscopic arrangement of kinesin-14 within a microtubule bundle remains unknown. The tight rectangular lattice suggests a lack of space for the kinesin motors between microtubules. One possibility is that kinesin is excluded from the bundle interior and can only bind to and power the motion of surface-bound microtubules. Alternatively, kinesin binding within the bundle interior would require the formation of highly localized filament deformations. These questions, which require alternative experimental methods such as electron microscopy, must be addressed before a complete multiscale understanding of microtubule-based active matter emerges. 

\begin{acknowledgments}
We thank the Streichan Lab for lending us time on their Leica SP8. Thanks to Wen Yan and Sebastian F\"{u}erthauer for their valuable discussions. We thank Youli Li and Phillip Kohl for their support in using the SAXS facilities at UCSB. We thank Uri Raviv and Asaf Shemesh for their help using the X+ and D+ packages. DJN acknowledges the support of NSF-DMR-2004380, NSF-DMR-1420570, and NSF-DMR-0820484. ZD acknowledges the support of NSF-DMR-2004617 and NSF-MRSEC-2011846. CEW and SEM acknowledge the support of NIH R35-GM122482. We also acknowledge the use of Brandeis MRSEC optical microscopy and biosynthesis facilities, funded by NSF-MRSEC-2011846.
\end{acknowledgments}

\bibliography{ms}% Produces the bibliography via BibTeX.

\clearpage

\onecolumngrid
\begin{center}
  \textbf{\large Supplementary Material}\\[.2cm]
\end{center}

\setcounter{equation}{0}
\setcounter{figure}{0}
\setcounter{table}{0}
\setcounter{page}{1}
\setcounter{section}{0}
\renewcommand{\theequation}{S\arabic{equation}}
\renewcommand{\thefigure}{S\arabic{figure}}

\section{\label{sec:SuppInfo} Supplementary Information}
 
\subsection{\label{sec:Methods} Methods}
\subsubsection{\label{sec:suppSamplePrep} Sample Preparation}
The full-length \textit{Xenopus} kinesin-14 clone XCTK2 was expressed in and purified from baculovirus-infected Sf-9 cells (Invitrogen). The active samples consist of GMPcPP (Guanosine-5'-[($\alpha$,$\beta$)-methyleno]triphosphate) stabilized microtubules mixed with kinesin-14 in a buffer of 80 mM PIPES (piperazine-N, N'-bis), 5 mM magnesium chloride,1 mM EGTA, with a pH of 6.8 adjusted with potassium hydroxide. Kinesin-14 hydrolyzes ATP (Adenosine triphosphate, 1.4 mM) into ADP (Adenosine diphosphate) to fuel chemical motion. In addition, we add 0.034\% pyruvate kinase (PK/LDH, Sigma P-0294) and 52 mM PEP (Phosphoenolpyruvate, Alfa Aesar B20358), which phosphorylates ADP (Adenosine diphosphate) into ATP to maintain constant kinesin dynamics. To prevent photobleaching, we added 4.2 mM DTT (dithiothreitol, ACROS Organics 16568), 2.5 mg/mL glucose (Sigma G7528), 0.03 mg/mL catalase (Sigma C40), and 0.17 mg/mL glucose oxidase (Sigma G2133). When noted, experiments include 35 kDa polyethylene glycol (PEG). 

Tubulin was purified from bovine brains \cite{castoldi2003}. It was polymerized and stabilized into microtubules by mixing 60 uM tubulin with 3 mM of the non-hydrolyzable GTP analog GMPcPP (Guanosine-5'-[($\alpha$,$\beta$)-methyleno]triphosphate, Jena Biosciences NU-405), and a solution of 1 mM DTT, 80 mM PIPES, 2 mM MgCl$_2$, 1 mM EGTA in DI water adjusted to a pH of 6.8 with KOH. In addition, we labeled 3\% of tubulin monomers with a fluorescent dye, Alexa-Fluor 647 (Invitrogen, A-20006), by a succinimidyl ester linker~\cite{hyman1991}. The solution was incubated in a water bath at 310 K for one hour and then left to cool to room temperature for 6 hours. Polymerized microtubules were flash-frozen in liquid nitrogen and subsequently thawed before creating a sample. To create tracer microtubules, we labeled tubulin with NHS-Azide and subsequently used Click chemistry to label microtubules with DBCO-488 \cite{Tayar2022}.

\subsubsection{\label{sec:suppChamberPrep} Chamber Preparation}
Experimental chambers had dimensions of 1.5 mm $\times$ 0.1 mm $\times$ 18 mm unless noted otherwise. The chamber consisted of a glass top and bottom, with parafilm spacers sealed with NOA 81 UV Adhesive (Norland Products, 8101) at both ends. The glass was coated with a polyacrylamide brush to suppress proteins' adsorption onto the glass~\cite{lau2009}. To bond to the glass, we warmed the parafilm and glass to 338 K and pressed the parafilm onto the glass with the rounded end of a PRC tube. This process led to chambers that are 80-100 $\mu$m in height.

\subsubsection{\label{sec:suppMicroscopy} Microscopy}
We used a Nikon Ti2 inverted microscope base equipped with an sCMOS detector (Andor Zyla) to take images using either a 4x Nikon Plan Apo Lambda (NA, 0.2) objective or a 40x Nikon Apo long working distance water immersion objective (NA, 1.15). Zeiss Immersol W, an NA-matched oil substitute, prevented imaging deterioration due to water evaporation during long acquisitions. We used a Leica SP8 Confocal with a 20x NA 0.75 air objective for confocal imaging and photobleaching experiments. We used a 633 nm laser to bleach the Alexa-647 labeled microtubules and simultaneously imaged unbleached Azide-DBCO-488 labeled tracer microtubules with a 488 nm laser. When using the Leica SP8 laser, it was essential to maximize bleaching efficiency, given the limitations of the laser power. Bleaching is most efficient at low magnification and high NA as $I \approx N_A^4/M^2$. To bleach bundles, we used a 633 nm laser on a Leica confocal to scan two thin lines \~15 $\mu$m apart for 5 seconds. The short duration of the bleached signal minimized the initial blurring of the bleach mark due to microtubule movement.

\subsubsection{\label{sec:suppSAXSExperiments} Small Angle X-Ray Scattering Experiments} 
We used a XENOCS Genix 50W x-ray microsource with a wavelength of 1.54 \AA and an EIGER R 1M detector. The sample to detector distance was 3.4 m, and images contain a Q range of $0.075$ nm$^{-1}$ to $2.185$ nm$^{-1}$. We used Silver Behenate as a calibration standard. We loaded samples into 1.5 mm Quartz Capillary Tubes (Hilgenberg  \#4017515) with 0.01 mm thick walls and then centrifuged the liquid to the bottom of the capillary by physically swinging the capillary tube. We sealed the capillaries using NOA 81 UV Adhesive (Norland Products, 8101). We aligned samples with the x-ray path near the bottom of the capillary. The detector integrated the X-ray scattering signal for 10 minutes, once an hour, for $\sim$7 hours unless otherwise noted. X-ray scattering data consists of two camera positions with an overlap. This multi-position acquisition allowed for the integration of signal at higher $q$. We masked and radially averaged the integrated scattering data using the Nika package for Igor Pro \cite{NikaForIgorPro}. The small-angle x-ray scattering (SAXS) data shown, unless otherwise stated, is the average scattering curve over 12 hours of acquisition. 

\begin{figure*}
\includegraphics[width=\textwidth]{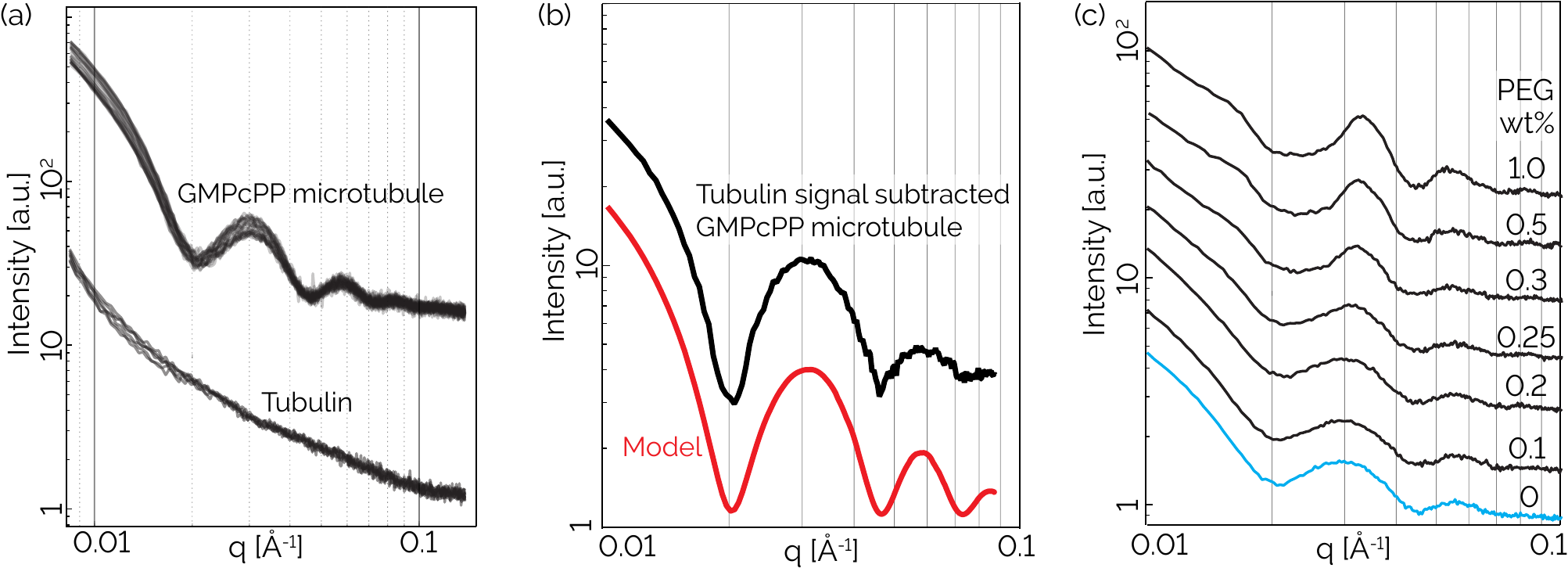}
\caption{ 
\textbf{Experimental scattering $I(q)$ of GMPcPP microtubules are modeled by a microtubule with $r_\textrm{MT}=$ 11.9 nm.}
\textbf{(a)} (top) The x-ray scattering curves from samples of GMPcPP polymerized microtubules without kinesin or PEG. (bottom) The x-ray scattering curve from samples of unpolymerized tubulin. 
\textbf{(b)} (top) The average x-ray scattering curve of GMPcPP polymerized microtubules after subtracting the measured unpolymerized tubulin curve. (bottom, red) A numerical scattering curve of a polymerized microtubule calculated by D+, with $r_\textrm{MT}=$ 11.9 nm.
\textbf{(c)} Radially averaged x-ray scattering curves of microtubules without kinesin with increasing concentrations of PEG. All curves shown have been arbitrarily shifted along the Y-axis.
}
\label{fig:supp_scattering} 
\end{figure*}

We loaded samples into capillary tubes and manually centrifuged the material to the bottom of the capillary tube. We sealed the tube with UV glue and imaged the capillary under a microscope to ensure the sample was active. We then loaded multiple capillary tubes onto a holder in front of a SAXS diffractometer and acquired scattering curves from each tube once per hour. $I(q)_{tubulin}$ of unpolymerized tubulin showed a decay with no resolvable peaks [Fig.~\ref{fig:supp_scattering}(a)]. $I(q)_{MT}$ due to polymerized microtubules displayed broad peaks at $ q \approx 0.028$ \AA$^{-1}$ and $ \approx 0.055$ \AA$^{-1}$. For instance $L=2\pi / (0.028$ \AA$^{-1}) \approx 22$ nm, which is roughly the diameter of a microtubule.

The experimental scattering $I(q)$ of microtubules after subtracting the background scattering due to free tubulin $I(q)_\textrm{tubulin}$ revealed the form factor $\hat{F}(q)$ of microtubules with no higher-order positional structure [Fig.~\ref{fig:supp_scattering}(b)]. We then performed SAXS experiments at various PEG concentrations with and without kinesin-14. We observed similar scattering curves at high PEG with and without kinesin but significantly different scattering curves at low PEG with and without kinesin [Fig.~\ref{fig:SAXS}(a), Fig.~\ref{fig:supp_scattering}(c)].
 
\subsection{\label{sec:SuppSAXS} Small Angle X-ray Scattering Modelling}

We used D+ to model the scattering amplitude of microtubule bundles. D+ is software that computes x-ray scattering curves of basic structures that are isotropically distributed \cite{DPlus}. First, we modeled a microtubule in the absence of PEG or kinesin. The model assumes a microtubule is a hollow cylinder with 13 protofilaments. Each protofilament is a rod with a diameter of 4.9 nm and a length of 2 $\mu$m. The microtubule's center-to-wall radius $r_\textrm{MT}$ is a free parameter. We compared the numerical model to the experimental scattering $I(q)$ of microtubules after subtracting the background scattering due to free tubulin $I(q)_\textrm{tubulin}$. Microtubules with a radius of $r_\textrm{MT}$ = 11.9 nm are consistent with the experiments [Fig.~\ref{fig:supp_scattering}(a,b)]. We used a solvent electron density of 333 electrons/nm$^3$ while the microtubule structures had an electron density of 411 electrons/nm$^3$.

We ran simulations of lattice configurations created by repeating patterns of $r_\textrm{MT}$ = 11.9 nm microtubules. We compared models to the scattering intensity $I(q)$ from kinesin-14 mediated microtubules in the absence of PEG, again subtracting the measured scattering $I(q)_{tubulin}$ due to free tubulin. We found that a hexagonal packing of ~6 microtubules with a spacing parameter of $L_h$ = 45.8 nm best fits the data. In contrast, other packing geometries, such as square or rectangular lattices, are incompatible with the experimental SAXS data at low PEG concentrations.

Previous SAXS experiments showed that at high PEG concentration microtubules have elliptical cross-sections and are arranged in a close-packed rectangular lattice \cite{dan2004, dan2005, dan2013}. Motivated by this information, we created simulations of microtubules with an elliptical cross-section. We calculated the scattering from microtubules arranged in close-packed rectangular lattices. The rectangular lattices consisted of 1-2 rows of microtubules along the short axis and 3-4 columns of microtubules along the long compressed axis. We included small (<1 nm) variations in microtubule position from an ideal rectangular lattice. The generated scattering curves from an ensemble of these lattices closely match the experimental scattering curves at high PEG concentrations. The rectangular packing has a spacing of $L_a$ = 13.6 nm and $L_b$ = 10.1 nm. The code to create these geometries is available at https://github.com/bezlemma. The broad shape of the peaks, and the correspondingly small number of microtubules in the lattice models, suggest that the microtubule packing is less crystalline and more liquid-like with short-range hexagonal or rectangular lattice patterns. 

\section{\label{sec:SuppFigures} Supplementary Figures}

\newpage

\section{\label{sec:SuppVideos} Supplementary Videos}

\begin{itemize}
\item \href{https://youtube.com/watch?v=-lbfLg1INY4}{Video 1}: The macroscopic dynamics of kinesin-14 driven microtubule bundles first without PEG and then with 1\% PEG. These dynamics appear qualitatively similar. In grey are fluorescent microtubules, and in yellow are individual tracer microtubules.
\item \href{https://youtube.com/watch?v=k6Z0WHoGG_s}{Video 2}: The motion of tracer microtubules (yellow) within a kinesin-14 driven microtubule bundle (not shown). Without PEG, the tracer microtubules display sliding and extending dynamics. In a sample with 1\% PEG, the dynamics of the tracer microtubules are still extending, but exhibit a stuttering motion.
\item \href{https://youtu.be/lTy-GO0qfzc}{Video 3}: A series of bleach line experiments between 0\% to 1\% PEG in several samples with differently sized bundles. In grey are fluorescent microtubules, which are bleached, and in yellow are individual tracer microtubules, which are not bleached. As PEG concentration increases, the samples transition to non-splitting dynamics. Between 0.2\% PEG to 0.5\% PEG, bleach lines reveal a coexistence between splitting and non-splitting dynamics.

\end{itemize}

\end{document}